\documentclass[conference]{IEEEtran}
\IEEEoverridecommandlockouts

\usepackage{cite}
\usepackage{amsmath,amssymb,amsfonts}
\usepackage{graphicx}
\usepackage{textcomp}
\usepackage{xcolor}
\usepackage{comment}
\usepackage{amsmath}
\usepackage{eqparbox}
\usepackage{cite}
\usepackage{amsmath,amssymb,amsfonts}
\usepackage{fixltx2e}

\usepackage{graphicx}
\usepackage{textcomp}
\usepackage{xcolor}
\usepackage{comment}
\usepackage{multirow}
\usepackage{subcaption}
\usepackage{balance}

\usepackage{algpseudocode}
\usepackage[utf8]{inputenc}
\usepackage{amsmath}
\usepackage{amsfonts}
\usepackage{amssymb}
\usepackage[ruled,vlined,linesnumbered]{algorithm2e}
\usepackage{hyperref}
\usepackage{algpseudocode}
\usepackage[linesnumbered,ruled,vlined]{algorithm2e}
\usepackage{nomencl}

\usepackage{fancyhdr}
\usepackage{lipsum} 
\fancypagestyle{accepted}{
    \fancyhf{}
    
    \fancyfoot[L]{\footnotesize \copyright 2024 IEEE. Personal use of this material is permitted. Permission from IEEE must be obtained for all other uses, in any current or future media, including reprinting/republishing this material for advertising or promotional purposes, creating new collective works, for resale or redistribution to servers or lists, or reuse of any copyrighted component of this work in other works. 
    Paper accepted at 35th IEEE International Conference on Application-specific Systems, Architectures and Processors (ASAP 2024).
    }
}

\def\BibTeX{{\rm B\kern-.05em{\sc i\kern-.025em b}\kern-.08em
    T\kern-.1667em\lower.7ex\hbox{E}\kern-.125emX}}
\begin{document}

\title{CHIME: Energy-Efficient STT-RAM-based Concurrent Hierarchical In-Memory Processing}

\author{
\IEEEauthorblockN{Dhruv Gajaria\IEEEauthorrefmark{1},
                     Tosiron Adegbija\IEEEauthorrefmark{1}, and
                     Kevin Gomez\IEEEauthorrefmark{2}}
    \IEEEauthorblockA{\IEEEauthorrefmark{1}Department of Electrical and Computer Engineering, University of Arizona, Tucson, AZ, USA\\
                     \IEEEauthorrefmark{2}Department of Computer Science, University of Massachusetts, Amherst, MA, USA\\
                     Email: \IEEEauthorrefmark{1}\{dhruvgajaria, tosiron\}@arizona.edu,
                           \IEEEauthorrefmark{2}kgomez@umass.edu}
}

\maketitle
\thispagestyle{accepted}
\begin{abstract}
Processing-in-cache (PiC) and Processing-in-memory (PiM) architectures, especially those utilizing bit-line computing, offer promising solutions to mitigate data movement bottlenecks within the memory hierarchy. While previous studies have explored the integration of compute units within individual memory levels, the complexity and potential overheads associated with these designs have often limited their capabilities. This paper introduces a novel PiC/PiM architecture, \textit{Concurrent Hierarchical In-Memory Processing (CHIME)}, which strategically incorporates heterogeneous compute units across multiple levels of the memory hierarchy. This design targets the efficient execution of diverse, domain-specific workloads by placing computations closest to the data where it optimizes performance, energy consumption, data movement costs, and area. CHIME employs STT-RAM due to its various advantages in PiC/PiM computing, such as high density, low leakage, and better resiliency to data corruption from activating multiple word lines. We demonstrate that CHIME enhances concurrency and improves compute unit utilization at each level of the memory hierarchy. We present strategies for exploring the design space, grouping, and placing the compute units across the memory hierarchy. Experiments reveal that, compared to the state-of-the-art bit-line computing approaches, CHIME achieves significant speedup and energy savings of 57.95\% and 78.23\% for various domain-specific workloads, while reducing the overheads associated with single-level compute designs.

\end{abstract}


\section{Introduction}
Data bottlenecks stemming from the memory hierarchy of contemporary computing systems remain a major performance-limiting factor, hindering the full potential of modern applications in von Neumann computing systems. This bottleneck arises from the increasing gap between processor speeds and the latency of data transfer across different memory hierarchy levels. As a result, processors often spend significant cycles idle, waiting for data, which hampers the performance of data-intensive workloads across domains like artificial intelligence, scientific computing, and real-time analytics. To address this issue, prior works have proposed the integration of compute units within the memory to enable processing in cache (PiC) and processing in memory (PiM) using bit-line computing \cite{gajaria2022study, jain2017computing,compute_caches} to mitigate the overhead associated with data transfers.

Bit-line computing is a computational approach that leverages the bit lines of memory architectures. By directly modulating the voltages or currents on these bit lines, computations on data (e.g., AND, OR, XOR operations) that share the same bit-line within a subarray can be achieved by activating multiple word lines. 
Previous research has predominantly focused on SRAM for Processing-in-Cache (PiC) and DRAM for Processing-in-Memory (PiM), driven by the premise that PiC and PiM may benefit different workloads in terms of energy and performance \cite{compute_caches,gajaria2022study}. However, deploying bit-line computing in SRAMs requires additional mechanisms to avoid data corruption from multiple word-line activations, increasing area overheads and memory access latencies \cite{jeloka201628}. DRAMs, meanwhile, require constant refreshing before bit-line computing operations, leading to energy overheads \cite{ankit2020circuits}.

STT-RAMs (Spin-Transfer Torque RAMs) have become an attractive option for bit-line computing, supporting both PiC and PiM operations. STT-RAMs have distinct read and write currents, which help mitigate data corruption issues, and their non-volatility eliminates the need for constant refreshing \cite{jain2017computing}. Previous work on bit-line computing with STT-RAMs primarily assessed the performance of primary compute operations at specific memory hierarchy levels \cite{gajaria2022study,jain2017computing}. The simplicity of these operations, mainly focusing on logical and addition operations, limits their real-world applicability, which often demands both simple and complex operations (e.g., multiplication). However, incorporating complex compute units for a broader range of operations in a single memory hierarchy level is impractical due to high area and latency costs.

\begin{figure}[t]
		\centering
		\includegraphics[width=0.85\linewidth]{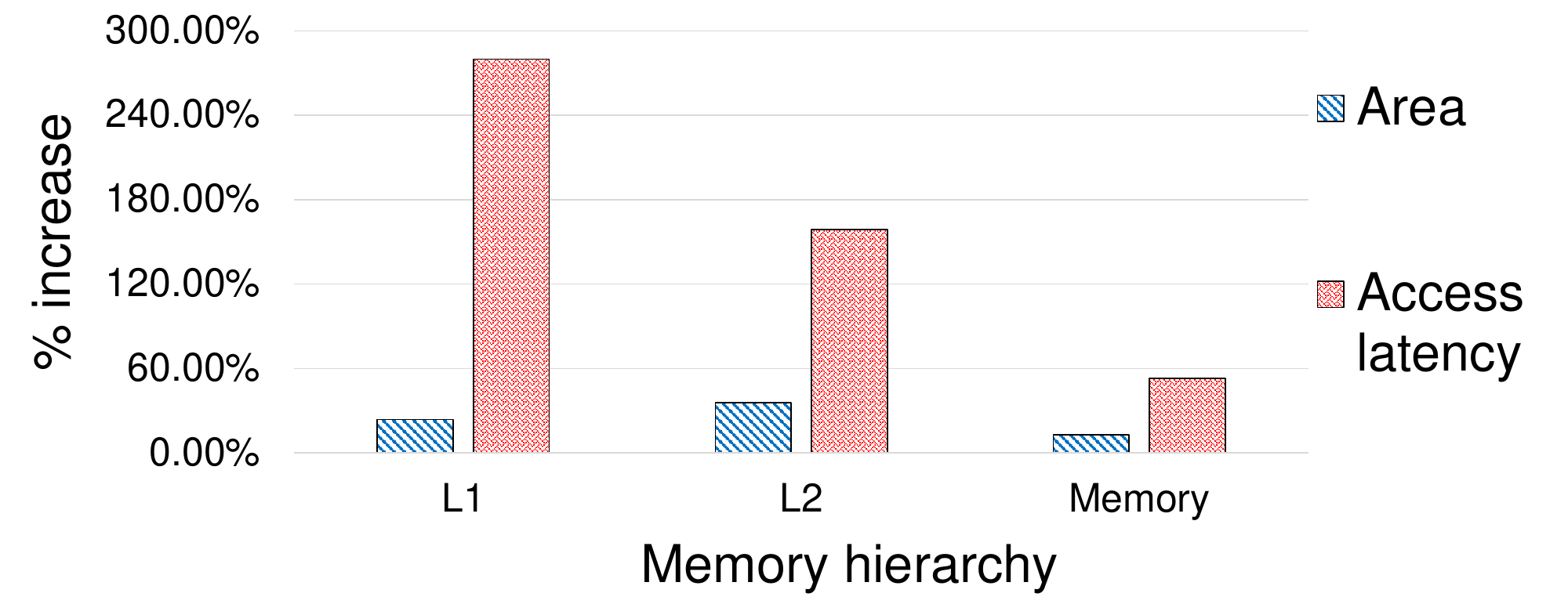}
		\caption{A compute unit with logical, add, subtract, shift, compare, and multiply operations adds significant overhead to a traditional cache/memory. }
		\label{fig:nvsim_intro}
\end{figure}

To demonstrate the challenges of deploying complex compute units in a single memory level, we performed experiments on L1 and L2 caches and main memory. These units featured operations, such as logical operations, shifts, additions, subtractions, multiplications, and comparisons. Our experimental setup is detailed in Section \ref{sec:exp_setup}. Our findings, illustrated in Figure \ref{fig:nvsim_intro}, show that these compute units increased the access latency and area by 2.7x and 25\% for the L1 cache, and 1.58x and 35.71\% for the L2. In the best case (with the lowest overhead), the latency and area increased by 53.02\% and 12.75\% for the main memory. These results highlight the substantial overheads associated with integrating complex compute units within traditional memory structures. This motivates us to introduce a novel approach called \textit{Concurrent Hierarchical In-Memory Processing (CHIME)}, which opportunistically distributes compute units across the memory hierarchy, enhancing computational concurrency while reducing overhead. As such, a broader range of operations can run in-cache or in-memory, while mitigating the overheads of provisioning a single memory hierarchy level with all the needed compute units.

To realize CHIME, we explore design-space exploration strategies for grouping compute units across memory hierarchy levels. Since compute units typically outnumber memory levels, we propose a domain-specific approach that creates \textit{compute groups} with heterogeneous compute units to match the memory levels, mitigating data transfer overheads across a range of workloads. We also explore strategies for \textit{mapping} these compute groups to specific memory hierarchy levels and demonstrate how CHIME enhances throughput by enabling concurrent bit-line computing at multiple memory hierarchy levels.

In summary, this paper makes the following important contributions:

\begin{itemize}
\item We introduce CHIME, a novel approach to hierarchical computing that utilizes heterogeneous compute units at various levels of the memory hierarchy tailored for complex bit-line computing operations across a range of operations in an application or domain.

\item We propose pipelined hierarchical computing to enable concurrency among compute units across different memory levels, facilitating faster program execution.

\item We present domain-specific design space exploration strategies for grouping and mapping diverse compute units to different memory hierarchy levels, given a set of workloads.

\item Our experimental results demonstrate significant improvements in latency and energy, 81.16x and 20.13x, respectively, compared to traditional CPU-based computing. CHIME also outperforms prior in-memory computing work, improving latency and energy by 57.95\% and 78.23\%, respectively, without introducing significant overheads.
\end{itemize}

\section{Background and related work}
STT-RAMs are gaining prominence as a memory technology due to their attractive combination of non-volatility, negligible leakage power, high density, endurance, and demonstrated commercial viability \cite{zhang2012multi}. At their core, STT-RAM cells consist of a magnetic tunnel junction (MTJ) and an access transistor. The MTJ comprises two ferromagnetic layers (a free and a fixed layer) separated by an oxide layer \cite{chun2012scaling}. Data is stored by altering the magnetization direction of the free layer relative to the fixed layer. However, this change in magnetization through spin-transfer torque incurs a trade-off: higher write latency and energy costs compared to traditional volatile memory technologies \cite{smullen2011relaxing}.

\subsection{Relaxed retention STT-RAM}
To mitigate STT-RAM write overheads, Smullen et al. \cite{smullen2011relaxing} proposed relaxing how long an STT-RAM cell can maintain its stored data (i.e., the \textit{retention time}). This can be achieved by altering the cell's magnetization saturation or effectiveness of the cell. We employ relaxed retention STT-RAM caches to mitigate the high write latency and energy as described in \cite{sun2011multi}. Kuan et al. \cite{kuan2019energy} explored STT-RAM caches in CPU-based computing, finding that different retention times optimize workloads based on their read-write ratios and cache block lifetimes. However, Gajaria et al. \cite{gajaria2022study} found that for bit-line computing on relaxed retention time caches, a homogeneous retention time can accommodate all workloads if the retention time exceeds the time required to load each cache line from the memory hierarchy's lowest level. Building on these insights, we use a homogeneous reduced retention time for our caches and non-volatile STT-RAM for main memory.

\subsection{Bit-line computing in STT-RAM}
Bit-line computing excels within non-volatile memories (NVMs) and SRAMs compared to DRAMs due to DRAM's inherent need for frequent refresh cycles \cite{ankit2020circuits}. This refresh requirement disrupts the computations performed directly on the bit-lines. To ensure reliable bit-line operations, a key requirement is the full recharging of source operand bit-lines. Previous studies have addressed this by explicitly refreshing the source operands prior to computation \cite{gao2019computedram}. This approach ensures consistent voltage levels on the bit-lines, enabling accurate in-memory calculations.

Unlike DRAMs, however, NVMs do not require constant data refreshes. Several prior studies have explored processing-in-memory (PiM) using NVMs \cite{jain2017computing, li2016pinatubo}. For instance, Jain et al. \cite{jain2017computing} proposed using bit-line computing with STT-RAM memories, featuring a compute unit that supports operations such as AND, OR, NOR, NAND, NOT, XOR, and ADD. 
Although Processing-in-Cache (PiC) has received relatively less attention than PiM, prior research \cite{compute_caches,eckert2018neural} has demonstrated its potential to mitigate data transfer overheads in specific workloads where PiM might not be as advantageous. 
Most related to our work, Gajaria et al. \cite{gajaria2022study} investigated STT-RAM-based bit-line computing at different memory hierarchy levels. Their findings suggest that the optimal placement for such computations depends on the workload characteristics. However, their work only focused on logical and add operations and did not cover other operations like multiplication, shift, comparator, etc. As such, operations not covered by the compute units were performed on the CPU, leading to a significant (up to 14x) reduction in the potential performance and energy optimization. Furthermore, while they only examined scenarios where all the compute units reside entirely within a single memory level, our work explores the benefits of opportunistically distributing a wide range of bit-line computing operations throughout different levels of the memory hierarchy.

\section{CHIME architecture}

\begin{figure}[t]
		\centering
		\includegraphics[width=0.85\linewidth]{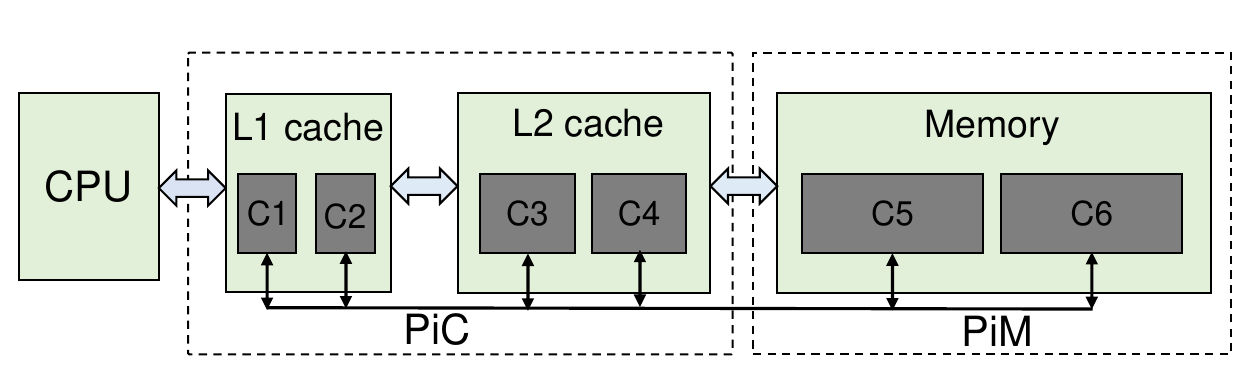}
		\caption{High-level overview of the proposed hierarchical in-memory processing approach, illustrating the flow of data and computation between low latency reduced retention STT-RAM caches (PiC) and non-volatile STT-RAM main memory (PiM). }
		\label{fig:system_model}
\end{figure}
While provisioning the main memory with all the in-memory compute units offers benefits, distributing compute units strategically across the memory hierarchy provides further advantages. Apart from enabling a higher throughput through concurrent hierarchical computing, this design recognizes the heterogeneous nature of workloads and their diverse computational requirements. This strategic placement not only balances performance with cost considerations but also significantly improves energy efficiency. Additionally, CHIME's distributed processing capabilities can exploit the transient nature of data blocks in some workloads. As data moves around the memory hierarchy, computations can be performed closer to the data's current location, further optimizing performance and energy consumption.

Figure \ref{fig:system_model} presents the CHIME system model, where the memory hierarchy levels (L1, L2, and main memory) contain different compute unit groups, catering to diverse compute demands. Unlike prior work, where unhandled complex computations are sent to the CPU, triggering significant performance loss and unnecessary data transfer \cite{gajaria2022study}, CHIME optimizes computations using bit-line computing across the memory hierarchy. Importantly, CHIME's hierarchical bit-line computing offers low access latencies (thanks to reduced retention caches) and high throughput, which helps offset the overheads associated with transferring data from main memory to the appropriate compute level (L2 or L1) when necessary. These benefits are reflected in our experimental results (Section \ref{sec:results}). Computations are only delegated to the CPU if the operation is unsupported within the compute units or the computation's characteristics (like highly sequential operations) would not gain significant performance advantages from in-memory or in-cache execution. This section briefly describes our bit-line computing architecture, compute unit designs, and strategy for enhancing concurrency throughout the memory hierarchy.

\subsection{Bit-line computing architecture}
\begin{figure}[t]
		\centering
		\includegraphics[width=0.99\linewidth]{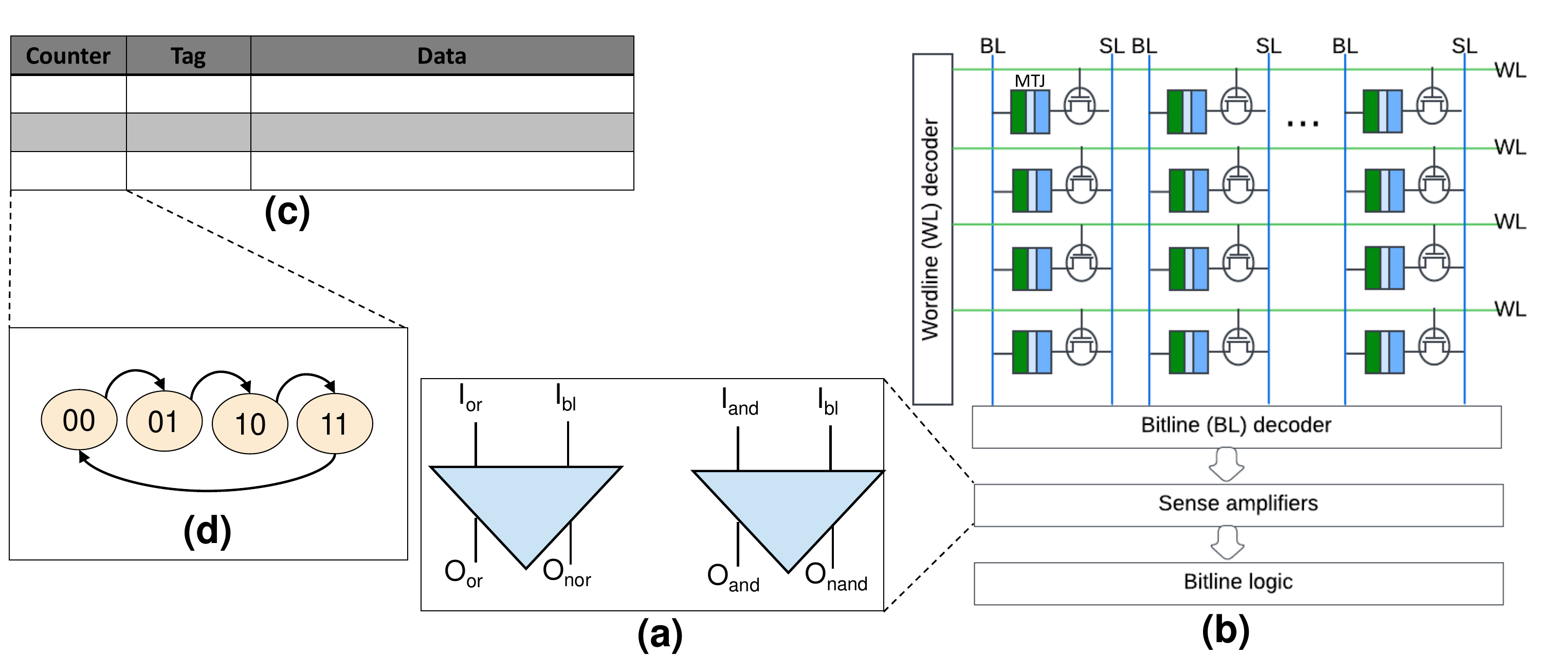}
		\caption{Illustrations of (a) modified sense amplifiers for bit-line computing, (b) STT-RAM cache subarray of MTJ cells with compute units after sense amplifiers, (c) components of a cache block with (d) a 2-bit cache monitor counter.}
		\label{fig:arcg}
\end{figure}

Figure \ref{fig:arcg} illustrates the bit-line computing architecture where multiple word lines within a subarray are activated simultaneously. The resulting bit-line charges ($I_{bl}$) are sensed by a specialized sense amplifier capable of performing logical operations, such as AND, OR, NAND, and NOR, directly on these signals. As shown in Figure \ref{fig:arcg}a, the second input to these computations is provided by reference currents ($I_{ref}$), such as $I_{and}$ and $I_{or}$, which are generated by dedicated circuits within the sense amplifier peripheral circuitry. These reference currents are programmed to specific values based on the desired logic operation and the expected bit-line current range. The output of the sense amplifier is a voltage signal (e.g., $O_{or}$, $O_{nor}$, $O_{and}$, or $O_{nand}$ in Figure \ref{fig:arcg}a) that reflects the result of the computation. The sense amplifier architecture is designed to compare the magnitudes of $I_{bl}$ and $I_{ref}$ to determine the output voltage levels ($O_{or}$, $O_{nor}$, $O_{and}$, or $O_{nand}$). For example, $I_{and}$ is set higher than the expected $I_{bl}$ when `0' is stored in any of the activated cells, ensuring the output ($O_{and}$) is low in that case. Conversely, $I_{or}$ is set lower than the expected $I_{bl}$ when a `1' is stored in any of the activated cells, ensuring the output ($O_{or}$) is high. 

Figure \ref{fig:arcg}b illustrates bit-line computing in a memory subarray. The subarray contains STT-RAM cells and the sense amplifier that provides inputs to the compute units (detailed in Section \ref{sec:design_compute_unit}). Figure \ref{fig:arcg}c depicts a memory block containing a tag and counter. We employ a counter (Figure \ref{fig:arcg}d) in relaxed retention caches to prevent data loss when the retention time elapses. The counter resets on each cache block write, and upon expiration, it invalidates the cache block and sends it to the lower memory level (if dirty), ensuring data preservation. The counter can be adjusted based on the retention time. For example, assuming an $N$-bit counter, where $N = 2$ with 4 states, and the retention time is 75$\mu$s, the counter clock is set to 18.75$\mu$s. In the caches, these counters have only a 0.75\% per-block area overhead. Counters are not used for non-volatile STT-RAM (the main memory) where data does not expire. 

While CHIME's architecture has a natural synergy with STT-RAMs, we note its potential compatibility with SRAMs after certain modifications. These modifications would address potential data corruption in SRAMs due to simultaneous word-line activation, likely involving low-voltage word lines \cite{jain2017computing,compute_caches,simon2020blade}. Additionally, SRAM-based CHIME would eliminate the need for retention counters (Fig \ref{fig:arcg}d) used in the STT-RAM version.  However, an evaluation of SRAM-based CHIME is outside the scope of this paper. For the focus of this paper, we concentrate on CHIME's implementation and benefits within STT-RAM-based systems.

\subsection{Target workloads}\label{sec: workloads}

\begin{table}
\caption{Target workloads in our implementations}
\centering
\begin{tabular}{|l|c|c|c|}
\hline
Category & Kernel & Input size  \\ \hline
Neural network & Binarized neural network (\textit{bnn}) & 1024*1024\\
Image processing & Image grayscale (\textit{img-grayscale}) & 1024*1024\\
Image processing & Image thresholding (\textit{img-thresholding}) & 1024*1024\\
Neural network & Multiply accumulate (\textit{mac}) & 1024*1024\\
Signal Processing & Matrix addition (\textit{mat\_add})& 1024*1024\\
Signal Processing & Matrix multiplication (\textit{mat\_mult})& 1024*1024\\
Signal Processing & Root mean square error (\textit{rmse})& 100000\\
Text Processing & String word count (\textit{wordcount})& 20000\\
\hline
\end{tabular}
\label{tab:workload}
\end{table}

The choice of supported compute operations and design space exploration in CHIME is fundamentally domain-specific and tailored to the characteristics of the target workloads. Understanding these workloads is crucial for effective customization. Therefore, we introduce our workloads here and use them as examples in the rest of the discussion. Similar to prior works \cite{gajaria2022study, jain2017computing, compute_caches}, we focus on bit-aligned array processing workloads. These workloads often exhibit abundant parallelism, allowing CHIME to fully leverage its distributed compute units. In addition, many core operations in these domains naturally align with bit-line computing, making implementation efficient. The workloads considered herein span signal processing, image processing, neural networks, and text processing---all areas where in-memory computing holds significant potential for acceleration. Table \ref{tab:workload} presents our workloads and their respective input sizes. The diversity in the workloads highlights the potential for CHIME's domain-specific customization across a range of important applications. 

\subsection{Designing various compute units}\label{sec:design_compute_unit}

We augmented the sense amplifier to support a range of operations, including AND, NAND, OR, NOR, and XOR (implemented using a combination of the basic gates). Multiplexers controlled by the cache control signals ensure flexible computation. The bitline logic block in Figure \ref{fig:arcg}b features additional compute units based on the workload needs. For instance, given the prevalence of comparisons in many of our target workloads, we incorporated a comparator for \textit{less than, greater than,} and \textit{equal to} operations. Additionally, a shift unit enables shift and rotate instructions, common in signal and image processing. For addition and subtraction, we opted for a ripple-carry design \cite{ripple_read} due to its reliability and suitability for our bit-line computing approach. For the multiplier, we chose a \textit{shift-and-add} \cite{shiftadd} multiplication design, among several options explored, to prioritize minimizing area and complexity. We used a 16-bit multiplier with 32-bit outputs, which provided sufficient precision for our target applications. All other compute units are designed to handle computations up to 32 bits, aligning with our workload data.

\subsection{Pipelined program execution}\label{sec:pipeline}

\begin{figure}[t]
	
		\centering
		\includegraphics[width=0.99\linewidth]{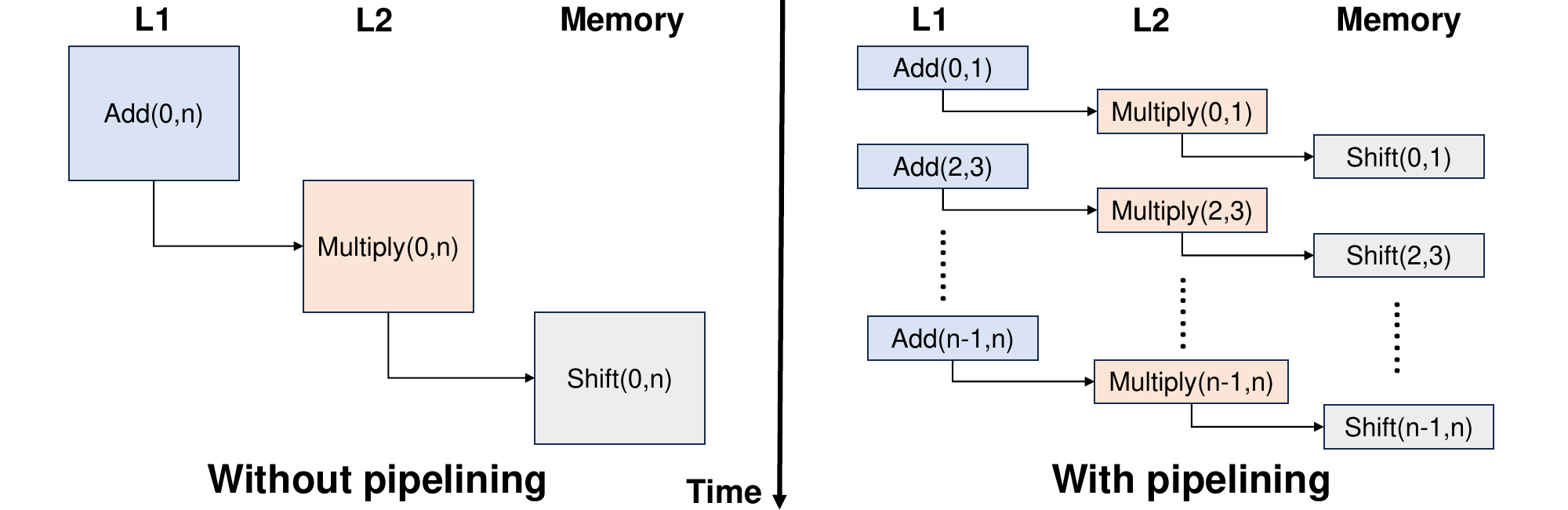}
		\caption{Program execution for hierarchical computing without and with pipelining.}
		\label{fig:cpu_dep}
\end{figure}

To enhance concurrency, CHIME employs an in-order pipelined execution flow for bit-line computing. Figure \ref{fig:cpu_dep} illustrates this concept, with the ADD compute units in the L1 cache, MULTIPLY units in the L2 cache, and SHIFT units in the main memory. Given a data flow from ADD to MULTIPLY operations, for example, rather than waiting for all ADD operations to finish, the completed results are moved to the L2 cache for concurrent execution with MULTIPLY operations, similar to vector chaining in SIMD computers. Besides enabling concurrency, this approach reduces latency by improving compute unit utilization. It also hides the data transfer latency across different levels of memory hierarchy since computations can happen within the memory while the data is being transferred. Since the target workloads exhibit data parallelism, the in-order execution ensures data readiness before executing the next instruction, thereby reducing stalls from data dependencies. Moreover, if pipelined operations in different levels access the same data simultaneously, the compiler ensures the data are available as needed. 
Compiler knowledge of the system enables workload profiling, operation scheduling, and data management, thereby obviating the need for runtime decision-making \cite{koehler2021towards}. While this limits our design to known systems, it is a realistic assumption for in-memory computing applications \cite{fujiki2019duality}.

We augment the compiler with simple instructions like $Instruction(Block_{a1}, Block_{a2}$, $Block_{dest})$. Here, $Instruction$ can be any supported operations ($ADD, SUB, OR$, etc.), and the addresses indicate where bit-line computations are performed. To control the data movement of specific cache blocks, $Move(Data, Hierarchy_{source}, Hierarchy_{dest})$ specifies the block address and the source and destination in the memory hierarchy. Additionally, $Copy(Data, Block_{dest})$ and $Invalidate(Data, Block_{a})$ will help create and invalidate copies of variables when necessary. These instructions guide the cache/memory controller in managing blocks for efficient bit-line computing. 

For our exploration in this work, we assume CHIME is employed within a single-core system. This focus allows us to simplify the design by avoiding the complexities of a cache coherency protocol specifically tailored to CHIME's distributed in-memory compute model. Additionally, since CHIME leverages significant parallelism within the caches and memory, we observe significant performance gains even within a single-core context. In future work, we plan to extend and investigate the use of CHIME within multi-core systems.

We note that the slowest memory hierarchy level will constrict the overall speed of CHIME's pipelined execution. This bottleneck is influenced by the latency of individual compute units and the available parallelism within that level. Therefore, to optimize performance, strategic grouping and mapping of compute units across the hierarchy are crucial. In the next section, we discuss these considerations in detail. 

\section{Design space exploration}
Efficiently allocating compute units across the memory hierarchy is challenging. This involves \textit{grouping} operations into compute units and \textit{mapping} them to specific memory levels. With domain-specific workload awareness, efficient grouping and mapping can minimize data movement, enhance concurrency, and optimize the throughput of bit-line computing. Factors like memory level characteristics---such as L1 cache's lower latency but smaller capacity for compute units compared to larger, slower memory levels---must also be considered. Similarly, more complex operations with a ripple-carry design like addition, subtraction, and multiplication, as discussed in Section \ref{sec:design_compute_unit}, would be slower than others. In this section, we explore various compute unit grouping and mapping strategies.

\subsection{Compute unit grouping}\label{sec:grouping}

\begin{algorithm}[t!]
\footnotesize
\caption{Compute unit grouping}
\label{algorithm:management}
    \SetKwInOut{Input}{Data}
    \SetKwInOut{Output}{Result}
    \DontPrintSemicolon
    \Input{
    List of workloads
    $W = [w_1, w_2, ....w_n]$
    
    List of bit-line computing instructions for each workload 
    $T = [t_1, t_2, ..., t_n]$\
    
$m$ = number of levels of the memory hierarchy. 
    }
    \Output{Compute unit groups}


$PairCount = []$


\For{$workload$ in $ workload_{list}$}{
    $CreateInstructionPair(T, workload)$\;
    
    \For{$instruction_{pair}$ in $ workload$}{
        \If{$instruction_{pair}$ not in $ PairCount$}{
            $PairCount.append(instruction_{pair})$\;
        }
        $PairCount[instruction_{pair}] += 1$\;
    }
}

$ComputeGroups = []$\;
$group = 0$\;

\For{$SortedPair$ in $Sort(PairCount, descending)$}{
    $ComputeGroup[group].append(SortedPair)\;$
    $group += 1$\;
    \If{$group > m$}{
        $group = 0$\;
    }
}

\end{algorithm}

We focus on grouping strategies that minimize data transfer overheads. These overheads occur when the compute unit needed by an operation is in a different memory level than its prior operation. Consequently, data read/write operations at different memory levels could increase computation-to-communication overheads and energy consumption.

Our grouping strategy, shown in Algorithm \ref{algorithm:management}, starts by breaking down target workloads into their individual operations. The goal is to count the unique instruction pairs---a set of two consecutive dependent instructions---across the workloads (lines 1-3) for compute group allocation.  If an instruction pair appears for the first time, the algorithm creates a $PairCount$ entry (line 6) and increments it for subsequent occurrences (line 7). Lines 3-7 are repeated for each workload to track the instruction pair frequency. After processing all workloads, the algorithm sorts the final pair count in descending order (line 10), prioritizing frequently occurring pairs for compute group formation (line 11). Based on pair frequency, sorted instruction pairs are assigned to a unique compute group (lines 11-12). If compute groups exceed the number of memory hierarchy levels, they are reset to the initial compute group and assigned accordingly (lines 12-13). This strategy prioritizes frequent dependent instruction pairs in a unique group, thereby reducing data transfer overheads, mitigating the latency impact of the compute units, and improving concurrency. 
For our workloads (Section \ref{sec: workloads}), the algorithm formed \textit{log-sub} (logical \& subtraction operations), \textit{add-comp} (addition \& comparisons), and \textit{mult-shift} (multiplication \& shift/rotate). While different grouping strategies can be used as per design needs, our focus here is on minimizing data transfer overheads. 

\subsection{Mapping compute groups to memory levels}\label{sec:mapping}
\begin{figure}[t]
		\centering
		\includegraphics[width=0.95\linewidth]{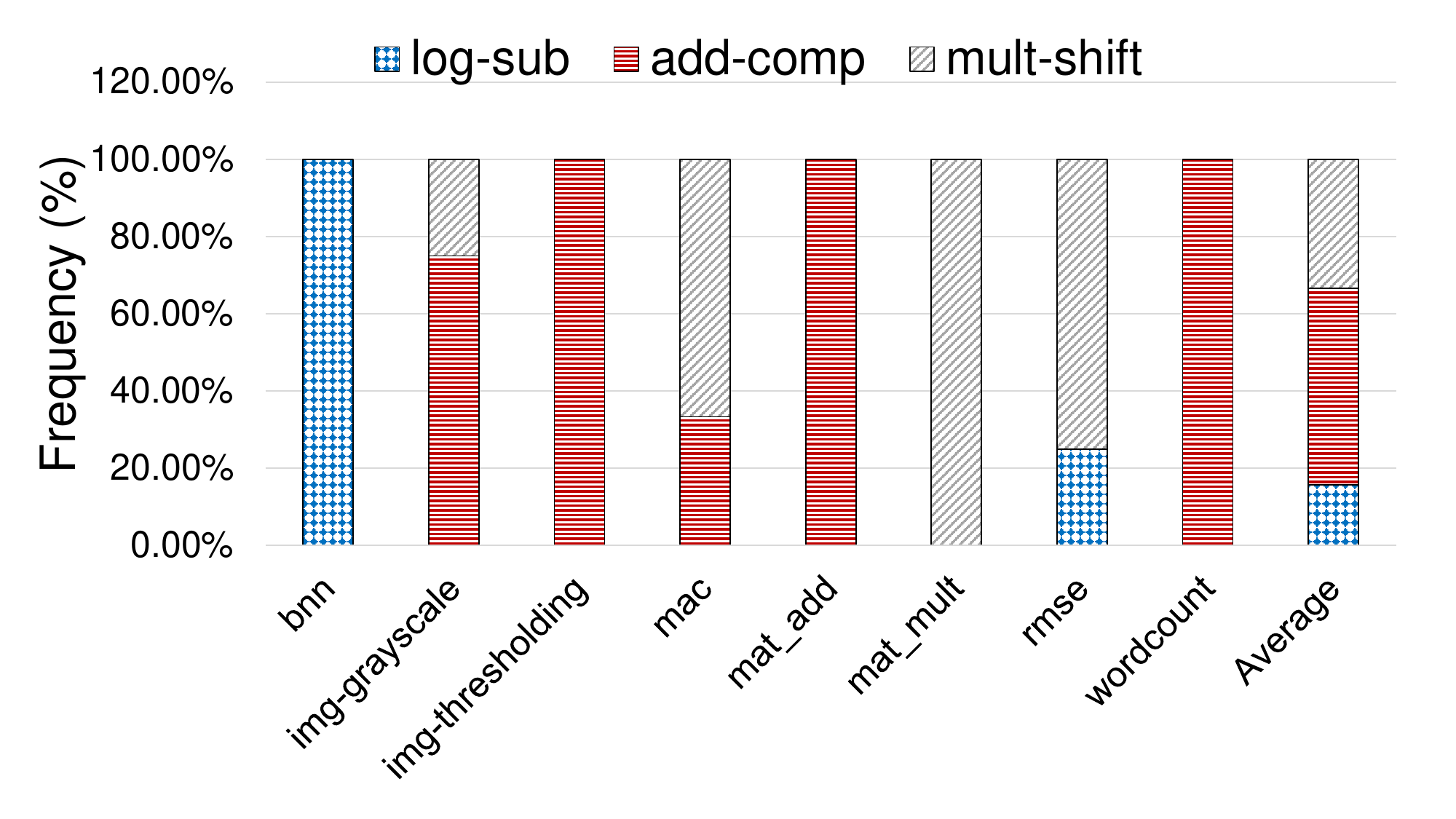}
		\caption{Frequency of compute groups for each workload.}
		\label{fig:inst_uti}
\end{figure}

Given the compute groups as determined by Algorithm \ref{algorithm:management}, we analyzed how frequently each compute group, comprising a set of dependent instruction pairs, is used with different workloads. Priority was given to the most frequently used compute groups. Figure \ref{fig:inst_uti} illustrates the frequency of the different compute groups for our workloads. Compute group \textit{add-comp} has the highest average frequency among all the workloads, followed by \textit{mult-shift} and then \textit{log-sub}. Therefore, this frequency dictates their mapping priority. 

We explored three mapping strategies for assigning compute groups to different memory hierarchy levels. The first strategy assigns the most frequently used groups to memory levels with the least resource constraints. The main memory has the least constraints and the L1 cache has the most due to their sizes. This strategy allows more compute units for frequently used groups, boosting their parallelism. 

As outlined in Section \ref{sec:pipeline}, CHIME's pipelined execution faces a throughput bottleneck at the memory hierarchy with the lowest performance. Therefore, our second mapping strategy prioritizes optimizing \textit{bit-line computing throughput} throughout the hierarchy. To achieve this, we first recalculate the compute latency for each compute group (obtained from Algorithm \ref{algorithm:management}). Next, we add the subarray access latency ($\text{subarray}_{\text{read + write}}$) for the relevant memory hierarchy level to determine the total compute unit latency ($\text{Latency}_{\text{total}}$) as shown in Equation \ref{eq:1}. The throughput for each group is then calculated by dividing the number of compute units by ($\text{Latency}_{\text{total}}$), as shown in Equation \ref{eq:2}. This strategy explicitly considers the trade-off between the higher latency and the greater abundance of compute units in lower memory levels. Our analysis found that this strategy yields the highest throughput in the L2 cache, followed by main memory, with the L1 cache having the lowest throughput. 

\begin{equation}
\begin{aligned}
\text{Latency}_{\text{total}} &= Latency(\text{compute}_{\text{group}}+\text{subarray}_{\text{read + write}})
\end{aligned}
\label{eq:1}
\end{equation}

\begin{equation}
\begin{aligned}
\text{Throughput} &= \frac{\text{Number of compute units}}{\text{Latency}_{\text{total}}} 
\end{aligned}
\label{eq:2}
\end{equation}


Our third mapping strategy specifically addresses the impact of ripple-carry operations on throughput. Recognizing that a simple 1-bit compute latency is insufficient for ADD, SUB, and MULTIPLY operations, we take ripple-carry propagation into account (Equation \ref{eq:1}, showing propagation delay for each operation as $\text{p}_{\text{operation}}$). First, we calculate the total carry propagation delay by multiplying the carry delay ($\text{Delay}_{\text{carry}}$) with the propagation distance (number of bits). Next, we determine the total ripple-carry latency ($\text{Latency}_{\text{ripple-carry}}$) by adding the total compute unit latency ($\text{Latency}_{\text{total}}$), as shown in Equation \ref{eq:4}. Finally, ripple-carry-aware throughput ($\text{Throughput}_{\text{ripple-carry}}$) is calculated by dividing the number of compute units by the total ripple-carry operation latency (Equation \ref{eq:5}). This approach seeks a balance between the number of compute units and their effective latency. Table \ref{tab:dse_mapping} presents the results of these three mapping strategies, each offering unique solutions, which are evaluated in the following section.

\begin{equation}
\text{Delay}_{\text{carry}} = 
\begin{cases} 
p_{\text{add}}, & \text{if ADD,} \\
p_{\text{sub}}, & \text{SUBTRACT, or} \\
p_{\text{mult}}, & \text{MULTIPLY} \\
0, & \text{otherwise}
\end{cases}
\label{eq:3}
\end{equation}

\begin{equation}
\begin{aligned}
\text{Latency}_{\text{ripple-carry}} &= (\text{Delay}_{\text{carry}} * \text{number of bits}) + \\
&\ \text{Latency}_{\text{total}}
\end{aligned}
\label{eq:4}
\end{equation}

\begin{equation}
\begin{aligned}
\text{Throughput}_{\text{ripple-carry}} &= \frac{\text{Number of compute units}}{\text{Latency}_{\text{ripple-carry}}} 
\end{aligned}
\label{eq:5}
\end{equation}

\begin{table}
\caption{Compute group mapping}
\centering
\begin{tabular}{|l|c|c|c|c|}
\hline
Strategy & L1 & L2 & Memory  \\ \hline
Number of compute units &\textit{log-sub} & \textit{mult-shift} & \textit{add-comp} \\
Throughput &\textit{mult-shift} & \textit{add-comp} & \textit{log-sub} \\
Ripple-carry-aware throughput&\textit{log-sub} & \textit{add-comp} & \textit{mult-shift} \\
\hline

\end{tabular}
\label{tab:dse_mapping}
\end{table}

\section{Experimental setup}\label{sec:exp_setup}
\begin{table}[t]

\renewcommand{\arraystretch}{1.00}
\caption{Cache and memory configurations}
\label{tab:cycles}
\centering
\vspace{-5pt}
\scalebox{0.85}{
\begin{tabular}{|c||c|c|c|}

    \hline
    Memory hierarchy				&L1 32kB-64B-4  &L2 1MB-64B-8 & Memory 8GB size\\
    \hline
   Retention time &75$\mu$s &10ms &5years \\
    \hline
    Total read latency &1 &2 &154 \\
    \hline
    Total write latency &2 &4 &110 \\
    \hline
   Subarray read latency&1 &2 &4  \\
    \hline
    Subarray write latency &2 &4 &5 \\
    \hline
    Read energy per-bit (pJ) &0.26 &0.88 &25.59 \\
    \hline
    Write energy per-bit (pJ) &3.30
 &6.13 &6.42 \\
    \hline
    Leakage power (mW)  &15.93   &281.63  &808.07 \\
    \hline
\end{tabular}}

\end{table}

We compared CHIME with traditional CPU-based computing and the most recent related work in STT-RAM bit-line computing with all the compute units deployed on one memory level \cite{gajaria2022study,jain2017computing}. We used a modified version of gem5 to simulate CPU-based computing and model the caches, employing a 2GHz ARM Cortex A72 CPU. We used the McPAT simulator \cite{li2009mcpat} to assess CPU power and energy consumption. SPICE simulations were used for compute units, and their outputs were passed to a modified NVSIM simulator \cite{dong2012nvsim} to model STT-RAM caches and memory with in-cache/memory compute units. 
Table \ref{tab:cycles} shows our cache configurations, latencies, and energies for read-write and bit-line computing operations. Lower memory levels require much larger subarrays, thereby increasing the energy and latency. 
Informed by workload analysis, we set retention times to 75$\mu$s for the L1 cache and 10ms for the L2 cache to minimize miss rates.  

\section{Results} \label{sec:results}
In this section, we begin by comparing the three compute group mapping strategies described in Section \ref{sec:mapping} to highlight the trade-offs and advantages of each approach. Subsequently, we compare CHIME with prior in-memory computing solutions to provide insights into CHIME's gains. Finally, we analyze the overheads associated with CHIME's implementation. To ensure a robust comparison, we implemented two variants of prior work (\cite{gajaria2022study}), with all the compute units in the L2 cache (\textit{STT-CiM\_L2}) or main memory (\textit{STT-CiM\_mem}). For CPU-based computing, the applications run on a single Cortex A72 CPU with energy-efficient STT-RAM caches enabling a rigorous comparison to CHIME. 
To ensure a fair comparison between CPU and bit-line computing, we only record statistics during the computation operations of the workloads and disregard input and output activities. 
In bit-line computing, data initially resides in the main memory, and data transfer overheads from main memory to caches are factored into our results.

\subsection{CHIME vs. CPU-based computing}\label{sec:res_mapping}

We compare the CPU to the three mapping strategies (Section \ref{sec:mapping}), i.e., based on the number of compute units (\textit{Compute}), throughput (\textit{Throughput}), and ripple-carry-aware throughput (\textit{RC-throughput}). 

\subsubsection{Latency}
Figure \ref{fig:dse_lat} shows the speedup of different mapping strategies compared to the base CPU. Compute, Throughput, and RC-throughput achieved average speedups of 45.75x, 37.41x, and 81.16x, respectively. The highest speedup was achieved for $wordcount$ at 95x, 295.2x, and 295.2x. CHIME's use of multiple compute units, reduced CPU dependency, and the parallelism achieved using bit-line computing contributed to this significant speedup. Among the mapping strategies, RC-throughput had the highest improvement, outperforming Compute and Throughput by 43.62\% and 53.90\%, respectively. 

RC-throughput achieved significant improvements for $mac$, $mat\_add$, $mat\_mult$, and $rmse$, as these workloads are dominated by ripple-carry ADD, SUBTRACT, and MULTIPLY operations. This shows the effectiveness of the RC-throughput in arranging the compute groups to achieve the best set of compute operations. Our compute grouping strategy successfully mitigated data transfer latencies, except in $bnn$, where a data transfer latency overhead of 4.36\% was incurred. $BNN$ had low data reuse, performed only bitwise $xor$ computations, and thus did not experience significant computation latency. 

\subsubsection{Energy}
\begin{figure}[t]
\centering
    \begin{subfigure}[t]{\linewidth}
      \centering
      \includegraphics[width=0.9\linewidth]{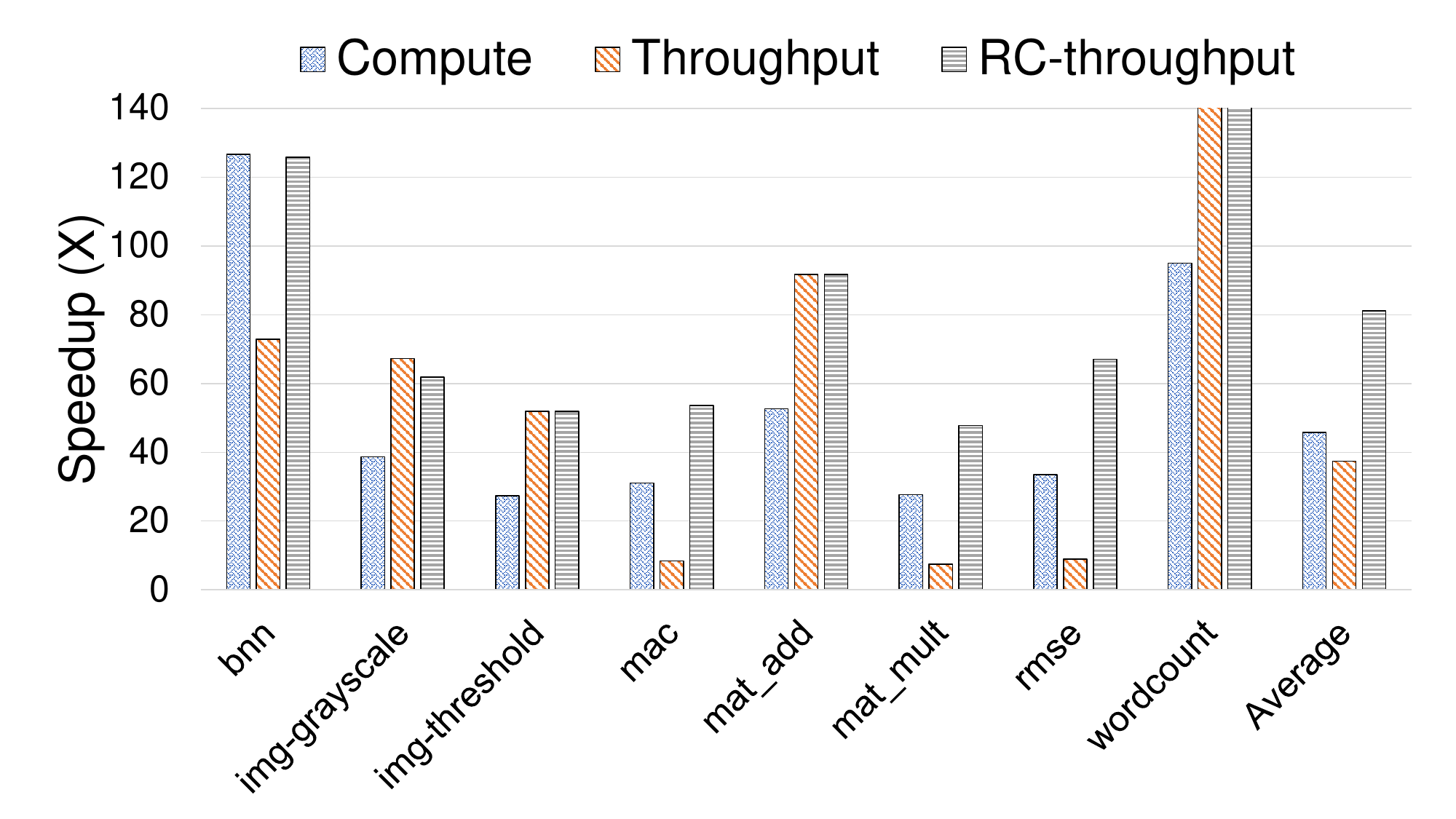}
      \caption{Latency}
      \label{fig:dse_lat}
    \end{subfigure}%

    \begin{subfigure}[t]{\linewidth}
      \centering
      \includegraphics[width=0.9\linewidth]{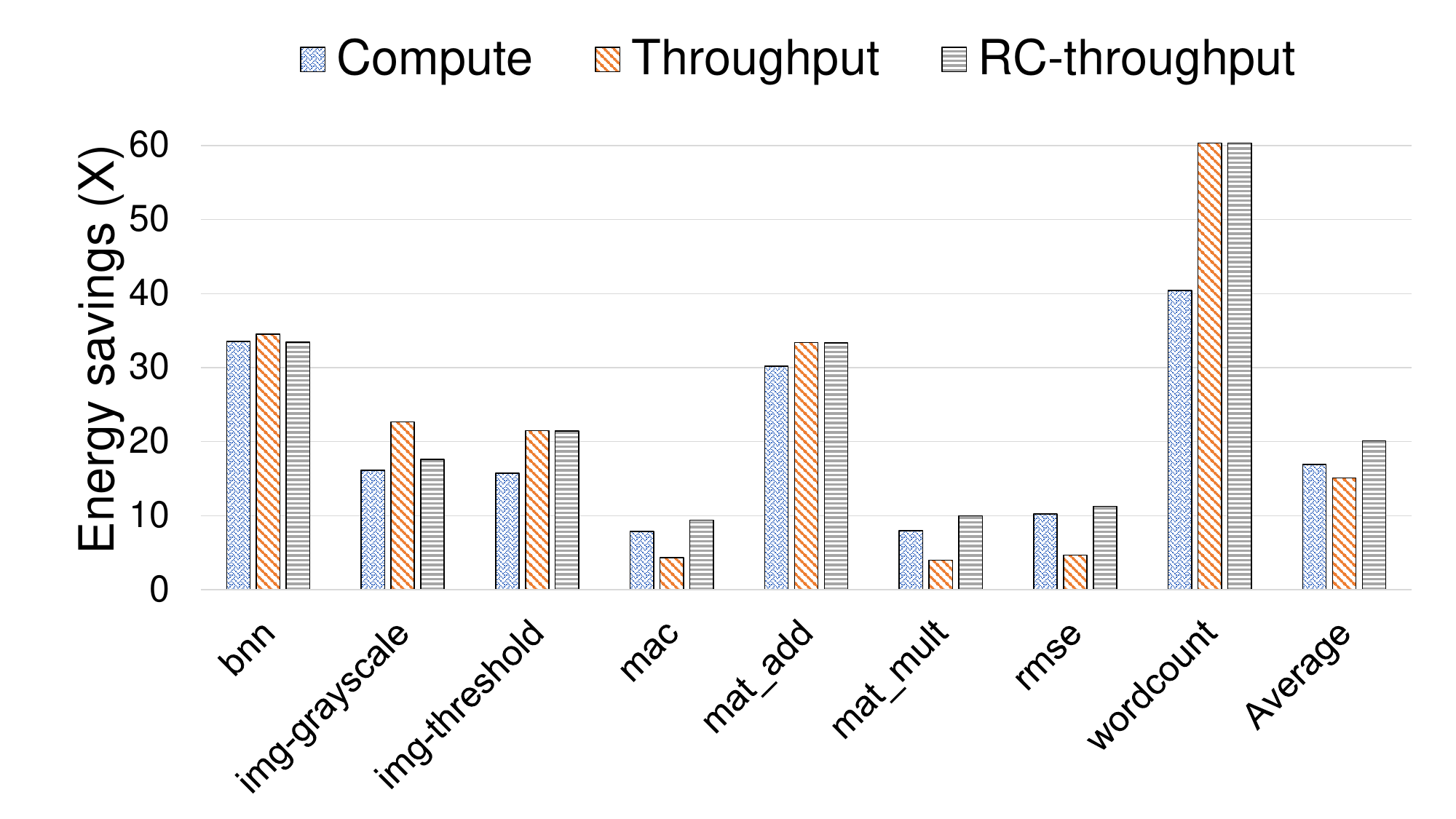}
      \caption{Energy}
      \label{fig:dse_enr}
    \end{subfigure}
 
\caption{Speedup and energy savings for different mapping strategies compared to CPU-based computing.}
\label{fig:dse_comp}
\end{figure}

Figure \ref{fig:dse_enr} shows the significant energy savings of different mapping strategies compared to the base CPU. On average, Compute, Throughput, and RC-throughput achieved energy savings of 16.92x, 15.13x, and 20.13x, respectively, with $wordcount$ having the highest savings of 40.43x, 60.32x, and 60.32x. RC-throughput outperformed Compute and Throughput by 15.93\% and 24.82\%, respectively. Due to our grouping strategies' focus on reducing data transfer overheads, we observed that compute and static energies dominated the total energy consumption. 

Although RC-throughput had the highest amount of data transfer energy (14.75\% of the total), it averaged 33.05\% and 52.2\% for static and compute energies, respectively. The data transfer energy for Compute and Throughput was at an average of 8.04\% and 8.23\%. However, Compute and Throughput had much higher static energy than RC-throughput, averaging 49.36\% and 55.63\%, respectively, due to lower performance. This shows the significant energy benefits of hiding the data transfer overheads behind the compute latency. 

\subsection{Comparison to the state-of-the-art}
To compare CHIME with the two variants (STT-CiM\_mem and STT-CiM\_L2) of prior work \cite{gajaria2022study}, we used RC-throughput as our mapping strategy, as it achieved the highest latency and energy optimizations. The compute unit for STT-CiM is complex, accommodating all compute operations within a single compute unit. 
Additionally, STT-CiM\_mem has a PiM-only architecture, similar to what is seen in \cite{jain2017computing}, while STT-CiM\_L2 has a PiC-only architecture, as seen in \cite{compute_caches}.  

\subsubsection{Latency}
Figure \ref{fig:pc_lat} presents the latency results for CHIME, STT-CiM\_mem, and STT-CiM\_L2 compared to the base CPU. CHIME outperformed the CPU with an 81.16x improvement, followed by STT-CiM\_mem and STT-CiM\_L2 with improvements of 35.68x and 34.12x, respectively. All our mapping strategies consistently outperform STT-CiM due to better concurrency afforded by CHIME. STT-CiM\_L2 only outperformed CHIME for $bnn$ (by 23.87\%) due to dominant data transfer latencies. However, on average, CHIME outperformed STT-CiM\_mem and STT-CiM\_L2 by 56.04\% and 57.95\%, respectively. 
STT-CiM\_mem had no data transfer overheads for all the applications, whereas STT-CiM\_L2 and CHIME could hide the data transfer overheads for all the applications except for $bnn$. CHIME had better compute performance than STT-CiM due to better throughput through CHIME's pipelined execution and efficient grouping and mapping strategies. Moreover, the high compute unit complexity in STT-CiM significantly increased cache and memory access latencies, affecting CPU operations. 

\subsubsection{Energy}
\begin{figure}[t]
\centering
    \begin{subfigure}[t]{\linewidth}
      \centering
      \includegraphics[width=0.9\linewidth]{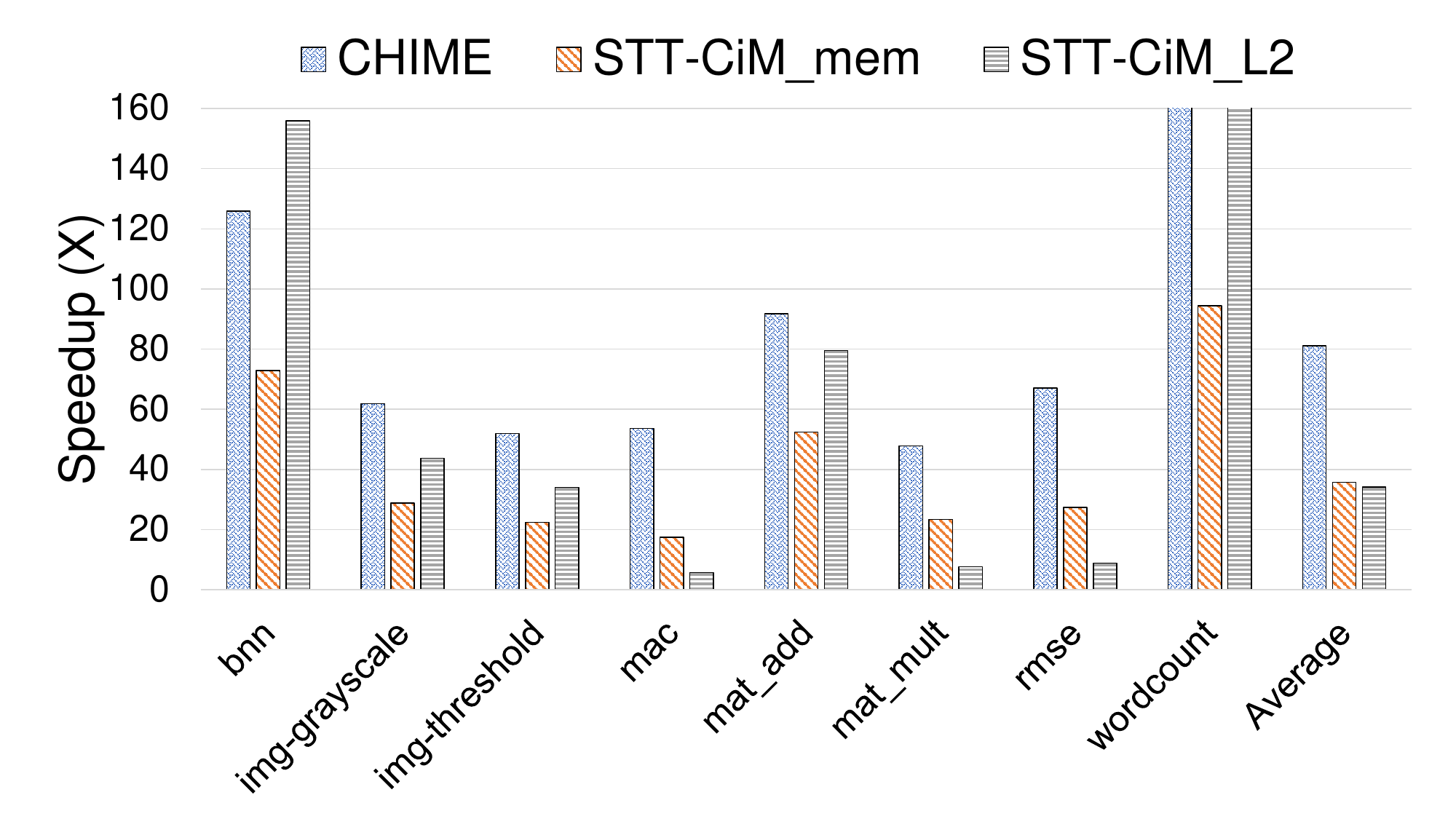}
      \caption{Latency}
      \label{fig:pc_lat}
    \end{subfigure}%

    \begin{subfigure}[t]{\linewidth}
      \centering
      \includegraphics[width=0.9\linewidth]{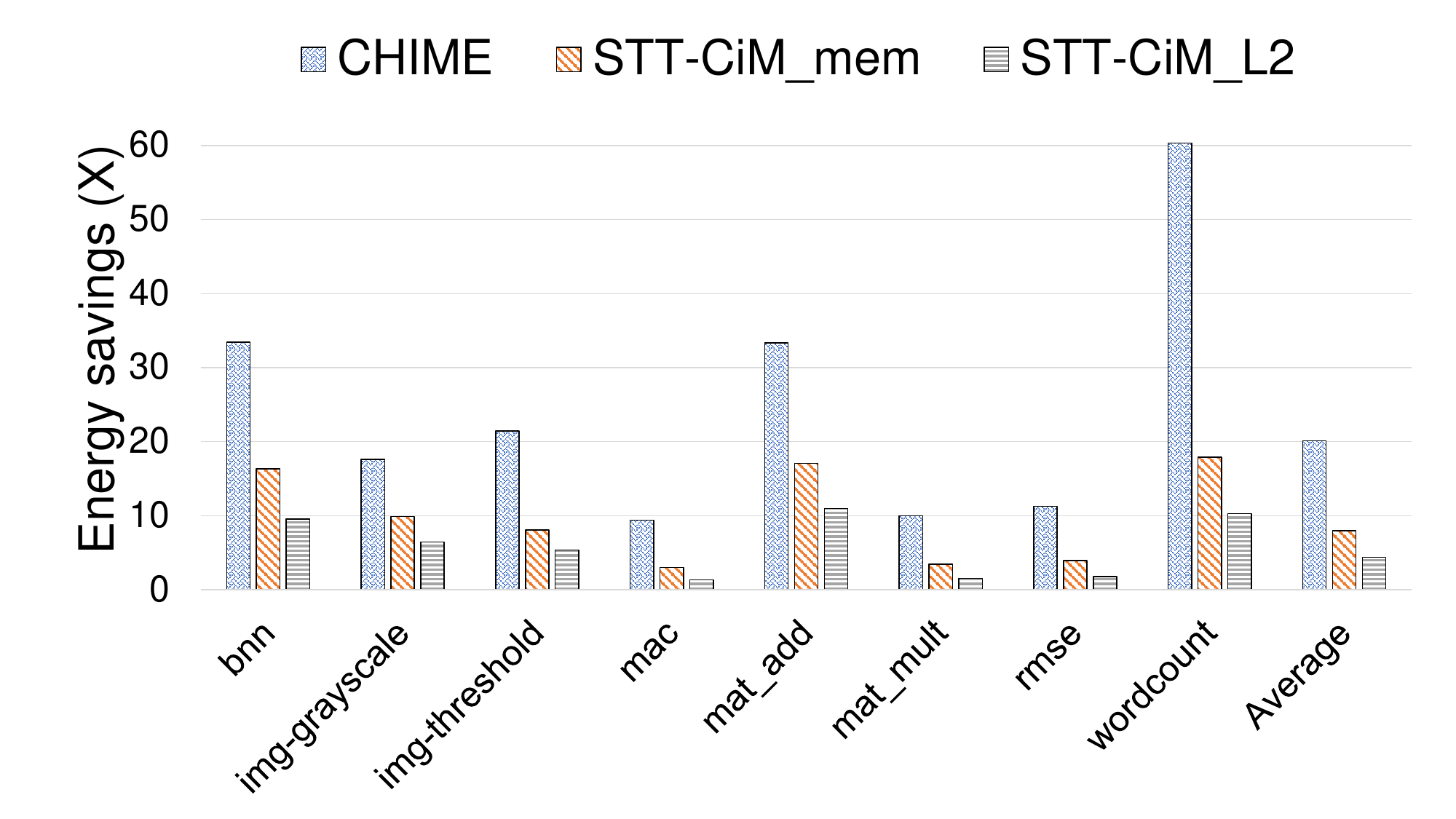}
      \caption{Energy}
      \label{fig:pc_enr}
    \end{subfigure}
 
\caption{Speedup and energy savings of CHIME and prior work (STT-CiM\_mem and STT-CiM\_L2) compared to CPU.}
\label{fig:pc}
\end{figure}

Figure \ref{fig:pc_enr} shows that CHIME similarly outperformed STT-CiM\_mem and STT-CiM\_L2 in energy savings compared to the base CPU. CHIME achieved a 20.12x energy improvement vs. STT-CiM\_mem and STT-CiM\_L2, which achieved improvements of 7.98x and 4.38x, respectively. Similarly to latency, CHIME achieved the highest energy savings for $wordcount$ (60.317x vs. 17.92x and 10.31x for STT-CiM\_mem and STT-CiM\_L2, respectively). On average, CHIME outperformed STT-CiM\_mem and STT-CiM\_L2 in energy savings by 60.35\% and 78.23\%, respectively. 

STT-CiM\_mem did not incur any data transfer energy costs, whereas the data transfer energy in STT-CiM\_L2 and CHIME were 5.11\% and 14.75\% of the total energy, respectively. However, the static energy for CHIME, STT-CiM\_mem, and STT-CiM\_L2 were respectively 52.20\%, 76.11\%, and 76.11\% of the total energy. This is because the pipelined execution of CHIME was more efficient than the STT-CiM\_mem and STT-CiM\_L2, which had to wait for compute elements to be idle to perform different computations, resulting in higher energy benefits in CHIME.
Overall, all CHIME mapping strategies outperformed STT-CiM, as STT-CiM's complex compute units incurred higher subarray overheads, increasing bit-line computing costs and exacerbating read-write operation overheads. 

\subsection{Overhead (critical path and area)}\label{sec:overheads}
For brevity, we report compute unit group overheads with respect to a 512*512 memory array. The compute groups' critical paths were 294ps, 265ps, and 452ps for \textit{log-sub}, \textit{add-comp}, and \textit{mult-shift}, respectively. These groups increased the static power by negligible amounts at 0.17\%, 0.20\%, and 0.21\%, respectively. CHIME introduced area overheads due to complex compute units and routing overheads, with \textit{log-sub} at 14.7\%, \textit{add-comp} at 17.12\%, and \textit{mult-shift} at 17.45\%. These overheads are far less than prior work STT-CiM's 25\% area overhead and 692ps critical path, demonstrating CHIME's effectiveness at achieving superior latency and energy despite lower overheads. Importantly, by facilitating bit-line computing for a wider variety of operations, CHIME prevents the optimization degradation because of CPU dependencies, as is the case in previous work \cite{gajaria2022study}.

\section{Conclusion}
This paper introduced \textit{CHIME}, a novel bit-line computing system model that distributes compute units across the memory hierarchy and enables concurrency to mitigate the effects of data transfer. The paper explored strategies for \textit{grouping} compute units and efficiently \textit{mapping} them to memory hierarchy levels. CHIME outperformed CPU-based computing with an 81.15x speedup and 20.12x energy savings, and surpassed the state-of-the-art by 57.95\% and 78.23\% in latency and energy savings. 
For future work, we aim to further investigate practical implementation strategies for a variety of real-world applications, specifically exploring the use of techniques like graph partitioning techniques to optimally group operations within the memory array. Additionally, we will develop intelligent mapping strategies to efficiently assign these operations to specific memory hierarchy levels to further minimize area overheads and improve overall throughput.

\section*{Acknowledgment}

This work was supported in part by the National Science Foundation under grant CNS-1844952. Any opinions, findings, and conclusions or recommendations expressed are those of the authors and do not necessarily reflect the views of the National Science Foundation.

\balance
\bibliographystyle{IEEEtran}
\bibliography{refs}


\end{document}